\documentclass[]{aastex631}

\newcommand{\roma}[1]{\uppercase\expandafter{\romannumeral#1}}
\newcommand{\speed}[1]{#1 km~s${}^{-1}$}

\newcommand{\acc}[1]{#1 m~s${}^{-2}$}
\newcommand{\nfig}[1]{Figure~\ref{#1}}

\newcommand{\mfig}[1]{Fig.\ref{#1}}

\usepackage{amsmath}
\usepackage{subfigure}
\usepackage{hyperref}%

 \usepackage{url}
\usepackage{graphicx}
\usepackage{natbib}
\usepackage{amssymb,txfonts}
\usepackage{multirow}
\usepackage{array}
\usepackage{sidecap}
\usepackage{hyperref}
\usepackage{epstopdf}
\usepackage{footnote}
\usepackage{tabularx}
\usepackage{booktabs}

\hypersetup{
   bookmarks=true,         % show bookmarks bar?
    unicode=false,          % non-Latin characters in Acrobat bookmarks
    pdftoolbar=true,        % show Acrobat toolbar?
    pdfmenubar=true,        % show Acrobat menu?
    pdffitwindow=false,     % window fit to page when opened
    pdfstartview={FitH},    % fits the width of the page to the window
    pdftitle={My title},    % title
    pdfauthor={Author},     % author
    pdfsubject={Subject},   % subject of the document
    pdfcreator={Creator},   % creator of the document
    pdfproducer={Producer}, % producer of the document
    pdfkeywords={keyword1} {key2} {key3}, % list of keywords
    pdfnewwindow=true,      % links in new PDF window
    colorlinks=true,        % false: boxed links; true: colored links
    linkcolor=blue,         % color of internal links (change box color with linkbordercolor)
    citecolor=blue,         % color of links to bibliography
    filecolor=blue%magenta,      % color of file links
    urlcolor=blue%cyan           % color of external links
    }

\shorttitle{Recurrent narrow quasi-periodic fast-propagating wave trains excited by the intermittent energy release in the accompanying solar flare}
\shortauthors{Zhou et al.}

\newcommand{\p}[1]{{\color{red}{#1}}}

\begin{document}

\title{Recurrent narrow { quasi-periodic fast-propagating} wave trains excited by the intermittent energy release in the accompanying solar flare}

\correspondingauthor{Xinping Zhou }
\email{xpzhou@sicnu.edu.cn}
\correspondingauthor{Yuandeng Shen}
\email{ydshen@ynao.ac.cn}

\author[0000-0001-9374-4380]{Xinping Zhou}
\affiliation{Sichuan Normal University, College of Physics and Electronic Engineering,  Chengdu 610068, People’s Republic of China}
\affiliation{Yunnan Observatories, Chinese Academy of Sciences,  Kunming 650216, People’s Republic of China}

\author[0000-0001-9493-4418]{Yuandeng Shen}
\affiliation{Yunnan Observatories, Chinese Academy of Sciences,  Kunming 650216, People’s Republic of China}

\author{ Hongfei Liang}
\affiliation{Yunnan Normal University, Department of Physics, Kunming 650500, People’s Republic of China}

\author[0000-0001-8318-8747]{ Zhining Qu}
\affiliation{Sichuan Normal University, College of Physics and Electronic Engineering,  Chengdu 610068, People’s Republic of China}

\author[0000-0001-9491-699X]{ Yadan Duan}
\affiliation{Yunnan Observatories, Chinese Academy of Sciences,  Kunming 650216, People’s Republic of China}

\author[0000-0003-0880-9616]{  Zehao Tang}
\affiliation{Yunnan Observatories, Chinese Academy of Sciences,  Kunming 650216, People’s Republic of China}

\author{ Chengrui Zhou}
\affiliation{Yunnan Observatories, Chinese Academy of Sciences,  Kunming 650216, People’s Republic of China}
\author{ Song Tan}
\affiliation{Yunnan Observatories, Chinese Academy of Sciences,  Kunming 650216, People’s Republic of China}

\begin{abstract}
About the driven mechanisms of the quasi-periodic fast-propagating (QFP) wave trains, there exist two dominant competing physical explanations: associated with the flaring energy release or attributed to the waveguide dispersion. Employing Solar Dynamics Observatory (SDO)/Atmospheric Imaging Assembly (AIA) 171 \AA\ images, we investigated a series of QFP wave trains composed of multiple wavefronts propagating along a loop system during the accompanying flare on 2011 November 11. The wave trains showed a high correlation in start time with the energy release of the accompanying flare. Measurements show that the wave trains’ phase speed is almost consistent with its group speed with a value of about \speed{1000}, indicating that the wave trains should not be dispersed waves. The period of the wave trains was the same as that of the oscillatory signal in X-ray emissions released by the flare. Thus we propose that the QFP wave trains were most likely triggered by the flare rather than by dispersion. We investigated the seismological application with the QFP waves and then obtained that the magnetic field strength of the waveguide was about { 10 Gauss}. Meanwhile, we also estimated that the energy flux of the wave trains was about $ 1.2\,\times 10^{5}\,  erg\cdot cm^{-2}{ s}^{-1}$.

\end{abstract}


\section{Introduction} \label{sec:intro}

Magnetohydrodynamic (MHD) wave theory in structured plasmas was formulated in the 1970s and 1980s \citep[e.g.,][]{1975IGAFS..37....3Z,1981SoPh...69...39R}. Over the past decades, modern observation instruments have captured various MHD wave modes \citep[see review by][]{2005LRSP....2....3N,2020ARA&A..58..441N,2020SSRv..216..140V}, such as standing fast kink waves \citep[e.g.,][]{ 1999ApJ...520..880A,1999Sci...285..862N,2002ESASP.506..461N,2008A&A...482L...9O}, global extreme-ultraviolet (EUV) waves \citep[e.g.,][]{2002ApJ...572L..99C,2005ApJ...622.1202C,2011LRSP....8....1C,2012ApJ...745L...5C,2012ApJ...752L..23S,2012ApJ...746...13L,2015ApJS..219...36G,2019ApJ...871....8L,2019ApJ...878..106H,2020ChSB,2020ApJ...894..139Z, 2020ApJ...905..150Z,2022ApJ...929L...4Z,2022ApJ...931..162Z,chen2022}, and trapped fast sausage waves \citep[e.g.,][]{ 2003A&A...412L...7N,2009A&A...503..569I,2020SSRv..216..136L}, providing detailed study conditions. 
In addition to playing an important role in coronal heating, the measured parameters of the MHD waves provide a condition for diagnosing coronal physical parameters that are hard to measure directly, such as the magnetic field strength \citep{2001A&A...372L..53N,2014ApJ...786..151S,2019ApJ...882...90L,2020ARA&A..58..441N,2020Sci...369..694Y,2021ApJ...921...61L,2021ApJ...908L..37M} through { seismological diagnostics} first proposed by \cite{1970PASJ...22..341U} and \cite{1984ApJ...279..857R}.

Recently, the narrow QFP magnetosonic wave trains\footnote[1]{According to the classification by \cite{2022SoPh..297...20S}, the QFP wave trains could be divided into two distinct categories, namely, broad and narrow QFP wave trains, based on different physical characteristics, such as observation waveband, propagation direction, angular width, intensity, and amplitude. The wave trains in the current study belong to the narrow QFP wave train.} { were} observed propagating along the funnel-shaped loop with a speed of 1000\,--\,\speed{2000} \citep{2011ApJ...736L..13L,2012ApJ...753...53S} with AIA \citep{2012SoPh..275...17L} telescope onboard SDO \citep{2012SoPh..275....3P}. It should be noted that their speed may be higher, as they usually take the projected velocity in the plane of sky. Compared with the broad QFP wave trains, the narrow QFP wave trains usually exhibit multiple successive arc-shape with a small angle extent of 10$^\circ$\,--\,60$^\circ$ and a weak amplitude intensity of 1\%\,--\,5\% \citep{2020ChSB,2022SoPh..297...20S}.

Since the initial discovery of the narrow QFP wave trains, many observational and simulation studies have been performed to investigate their triggering mechanisms and physical nature. These wave trains were well reproduced and identified as fast-mode magnetosonic waves by \cite{2011ApJ...740L..33O}, using a three-dimensional MHD model of a bipolar active region structure. 
Generally, there are several prominent opinions on their generation mechanisms. In the first case, it is considered that the formation of a QFP wave train is due to the dispersive evolution of an { impulsive broadband perturbation}, and their period is determined by the physical parameters of the waveguide and its surrounding environment \citep{1983Natur.305..688R,2004MNRAS.349..705N,2018MNRAS.477L...6S}. In the second case, they are believed to be triggered by the pulsed energy release in magnetic reconnections, and their periodicity is basically determined by the quasi-periodic pulsations (QPPs) in the accompanying flares \citep{2007LNP...725..221N,2012ApJ...753...53S,2013A&A...554A.144Y,2016ApJ...823..150T,2017ApJ...851...41Q,2018ApJ...859..154N,2018ApJ...868L..33L,2018ApJ...855...53L,2021SoPh..296..169Z,2022ApJ...930L...5Z,2022A&A...659A.164Z,2022ApJ...936L..12W}.  The relationship between the QPPs and the MHD waves can be found in review by \cite{2009SSRv..149..119N}. In addition, \cite{2013SoPh..288..585S} proposed that the leakage of pressure-driven photosphere oscillations to the corona should be one of the essential driving mechanisms. These divergences show that there is no unified explanation for the driving mechanism of the narrow QFP wave trains. Therefore, more observational studies are required to investigate their excitation mechanisms.

Although a few confident observations of narrow QFP waves have been reported in the literatures, adding new observations is valuable for understanding their physical properties and theoretical modeling. In this paper, we report a QFP wave trains that have a tight connection with the QPPs in the accompanying flare. The AIA images have a pixel size of 0.6 { arcsec and a cadence} of 12 seconds. The soft X-ray fluxes recorded by the Geostationary Operational Environmental Satellite (GOES) are used to study the periods of the associated flare. To study the effect of the disturbance on the plasma temperature and the density during the wave's passage, we calculate the Differential Emission Measure (DEM) for optically thin emission lines from plasma in the thermodynamics equilibrium. We perform seismology to estimate the magnetic field strength of the loop system. The primary analysis results are presented in Sect.\ref{se:results}. The discussion and the conclusion are given in Sect.\ref{se:discussion}.

\section{Results}
\label{se:results}
\subsection{Determination of the waves parameters}
\label{se:para}
\begin{figure}
\centering
\plotone{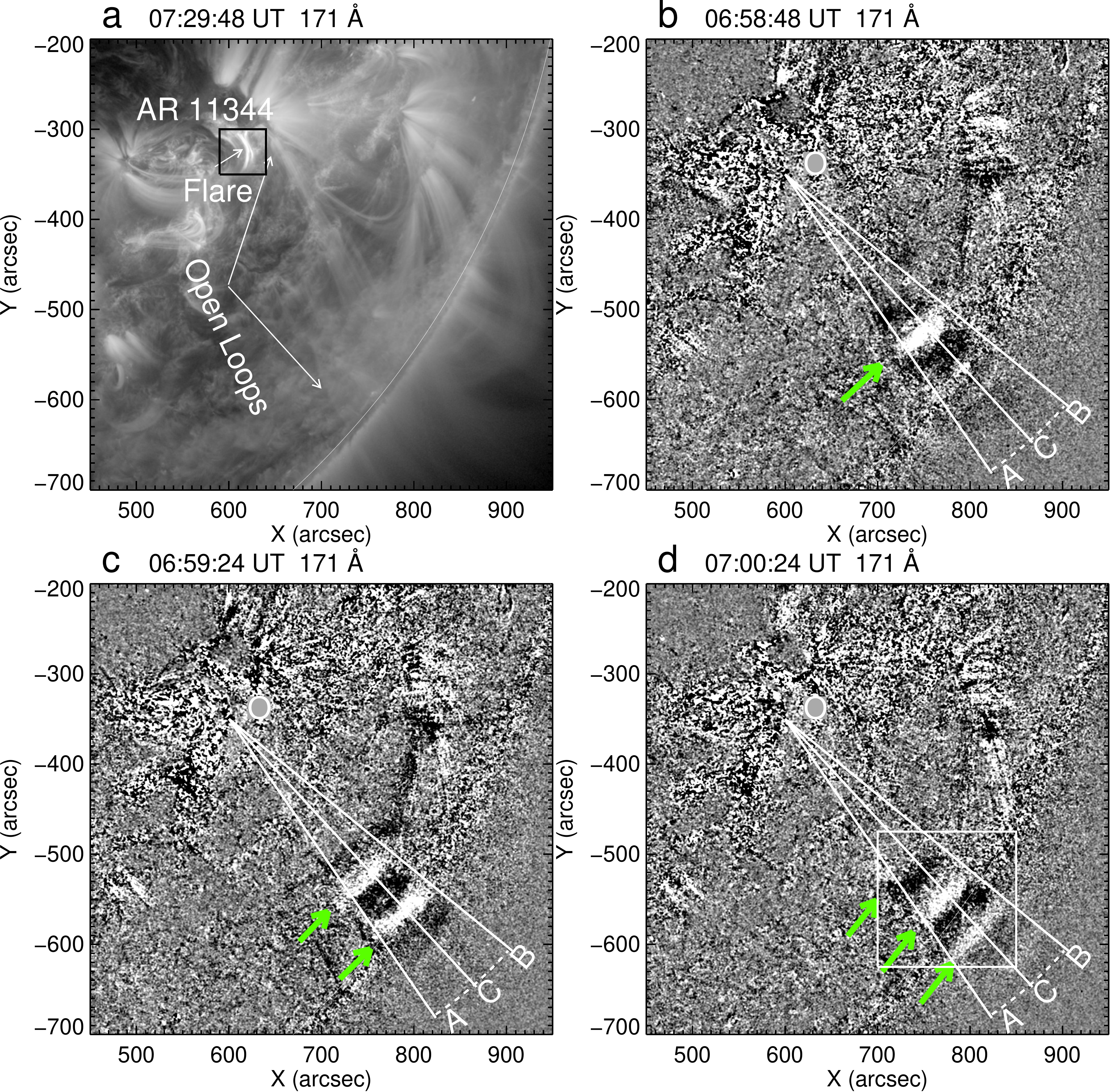}
\caption{Panel (a) shows the initial coronal condition of the eruption source region using the AIA/SDO 171 \AA\ direct images. The AIA/SDO 171 \AA\ running difference images in panels (b)-(c) show the snapshot of the QFP wave trains at three different times, in which the green arrows indicate the wave fronts. The { axis} OC of sector OAB (width 16$^\circ$) is used to obtain TDSP to trace the evolutions of the wave trains, and the resultant TDSP is shown in \mfig{fig:tdp} (a). The white box in panel (d) marks the region for Fourier analysis in Section \ref{se:para}. { An animation of panels (b)-(d) is available. The animation covers 06:30:24 UT-07:29:48 UT with a 12 seconds cadence. The animation duration is 10 seconds. }(An animation of this figure is available.)
\label{fig:overview}}
\end{figure}

The event occurred on 2011 November 11 in active region NOAA AR 11344. The waves were accompanied by a GOES C4.2-class flare\footnote[2]{\url{https://hesperia.gsfc.nasa.gov/goes/goes_event_listings/}}, whose start and peak times were 06:11UT and 07:05 UT, respectively. The pre-eruption coronal environment is displayed in \mfig{fig:overview} (a), in which a cluster of funnel-like open loops was rooted in NOAA AR 11344 and extending southwest of the disk limb. Three snapshots of 171 \AA\ running difference images in \mfig{fig:overview} (b)-(d) show the evolutions of the wave trains at 06:58:48 UT, 06:59:24 UT, and 07:00:24 UT, respectively, in which the green arrows mark the positions of the wavefronts. The sector OAB with angle 16$^\circ$ outlined the loop system where the QFP wave trains propagate. { These wave fronts are not observed in other EUV images, suggesting a narrow temperature response range. }

\begin{figure}
\centering
\plotone{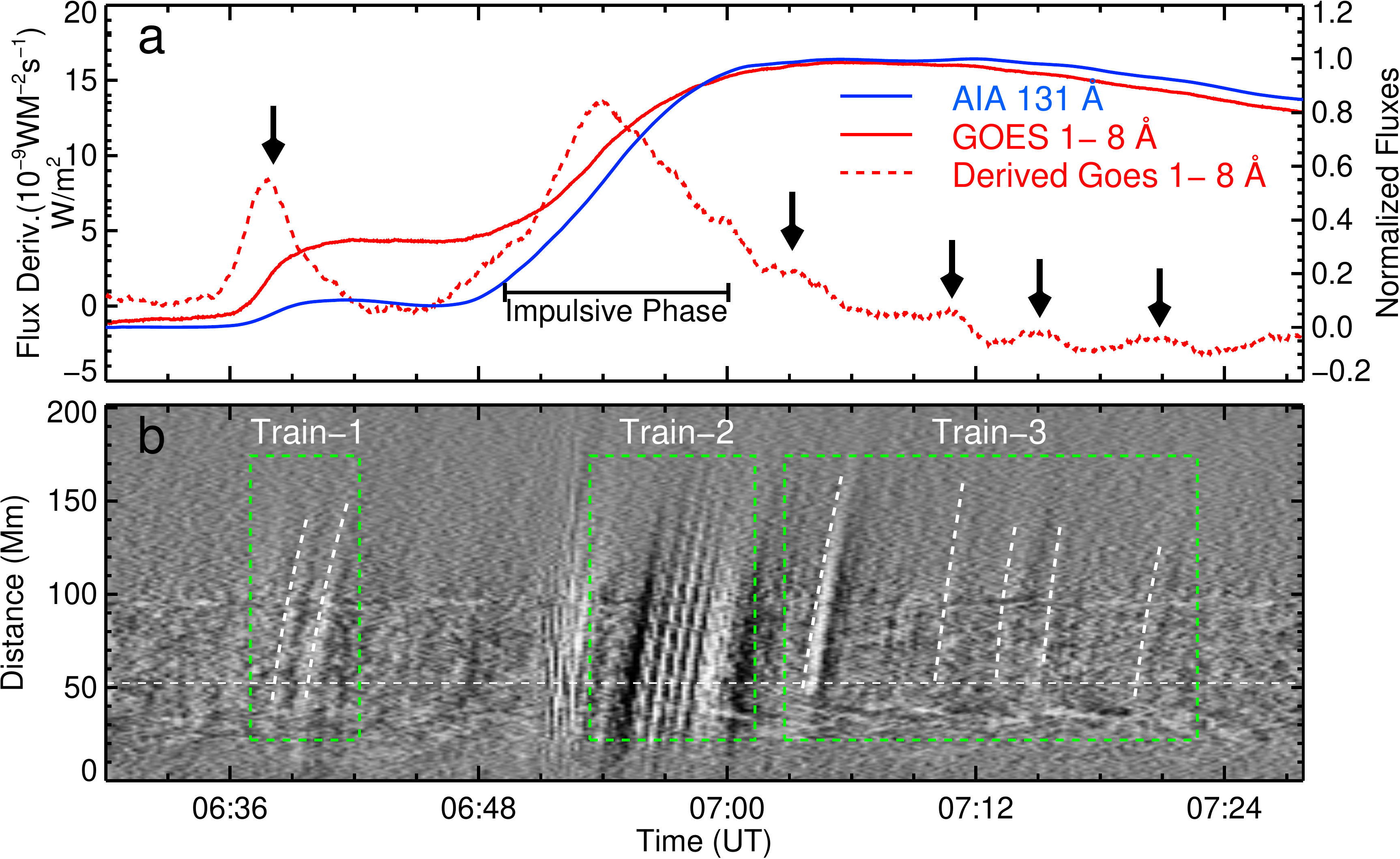}
\caption{Panel (a) displays the GOES 1\,--\,8 \AA\ X-ray flux curve, its { time derivative} curve, and normalized AIA 131 \AA\ light flux curve within the eruption source region outlined by the black box in \mfig{fig:overview} (a). Panel (b) is the TDSP obtained from the AIA 171 \AA\ running-difference images along the slice OC shown in \mfig{fig:overview}, in which the three green boxes labeled with Train-1, Train-2, and Train-3 point to the locations of the wave trains, while the white horizontal dashed line marks the position for extracting the signals for wavelet analysis in Section \ref{se:para}. \label{fig:tdp} }
\end{figure}

{ The QPPs are present in other EUV images. Here we only use the 131 \AA\ light curve measured from the flaring kernel marked with the black box in \mfig{fig:overview} (a) to obtain the information of flare.  }
\nfig{fig:tdp} (a) displays the GOES { 1-8} \AA\ flux curve (solid red curve) and the intensity of the 131 \AA\ light curve (solid blue curve). The similar trend of these two flux curves indicates no violent solar eruption in other active regions during this period. Thus, the GOES flux here can reflect the energy releases of the interested flare. Unfortunately, the Reuven Ramaty High-Energy Solar Spectroscopic Imager (RHESSI) did not observe during the flare. We use the time derivative of GOES soft X-ray 1-8 \AA\ flux as a proxy of the hard { X-ray} for the flare according to the Neupert effect \citep{1968ApJ...153L..59N,2005ApJ...621..482V,2010ApJ...717.1232N}. We derived the GOES 1-8 \AA\ flux's  { time derivative} and plotted the result with the red dotted line in \mfig{fig:tdp} (a). From the time derivative curve, we can identify several distinct bumps, pointed with black arrows, appearing in front and behind the impulsive phase. Generally, these bumps' appearance may manifest the flare's periodic energy release process.

To analyze the evolution and kinematics of the wave trains, we made the time-distance stack plot (TDSP) using the AIA 171 \AA\ running-difference images along the { axis} OC of the sector OAB (see \mfig{fig:overview} (b)-(d)). The result is shown in \mfig{fig:tdp} (b), where one can find that the wave trains could be distinguished as three wave trains, i.e., Train-1, Train-2, and Train-3. The signals of waves Train-1 and Train-3 are depicted by white slanted lines to highlight the boundary of wavefronts in the TDSP.
 Combining \mfig{fig:tdp} and the animation of \mfig{fig:overview}, we can identify that the wave trains are correlated with the flaring energy releasing: The first bump arose at about 06:36 UT, followed immediately by the occurrence of wave Train-1 with weak intensity; During the impulsive phase stage 06:48 UT-07:00 UT, representing the most energetic stage of energy releasing, Train-2, with multiple consecutive distinct wavefronts, emanated from the active region; Although the signals of wave Train-3 were fragile (except the first wavefront), we can also find that the start times of its wavefronts essentially were the same as that of the bumps during the delay phase stage. Therefore, we infer that the energy released by the flare excited the QFP wave trains. The { fluctuation} of the time derivative curve is a manifestation of the energy-releasing process below the flare due to the magnetic reconnection \citep{2015ApJ...807...72L,2015ApJ...813...59L,2022ApJ...931L..28L}. Thus the wave trains and { time derivative} curve of the GOES 1-8 \AA\ flux exhibit high correlations in start times as excepted.

\begin{figure}
\centering
\plotone{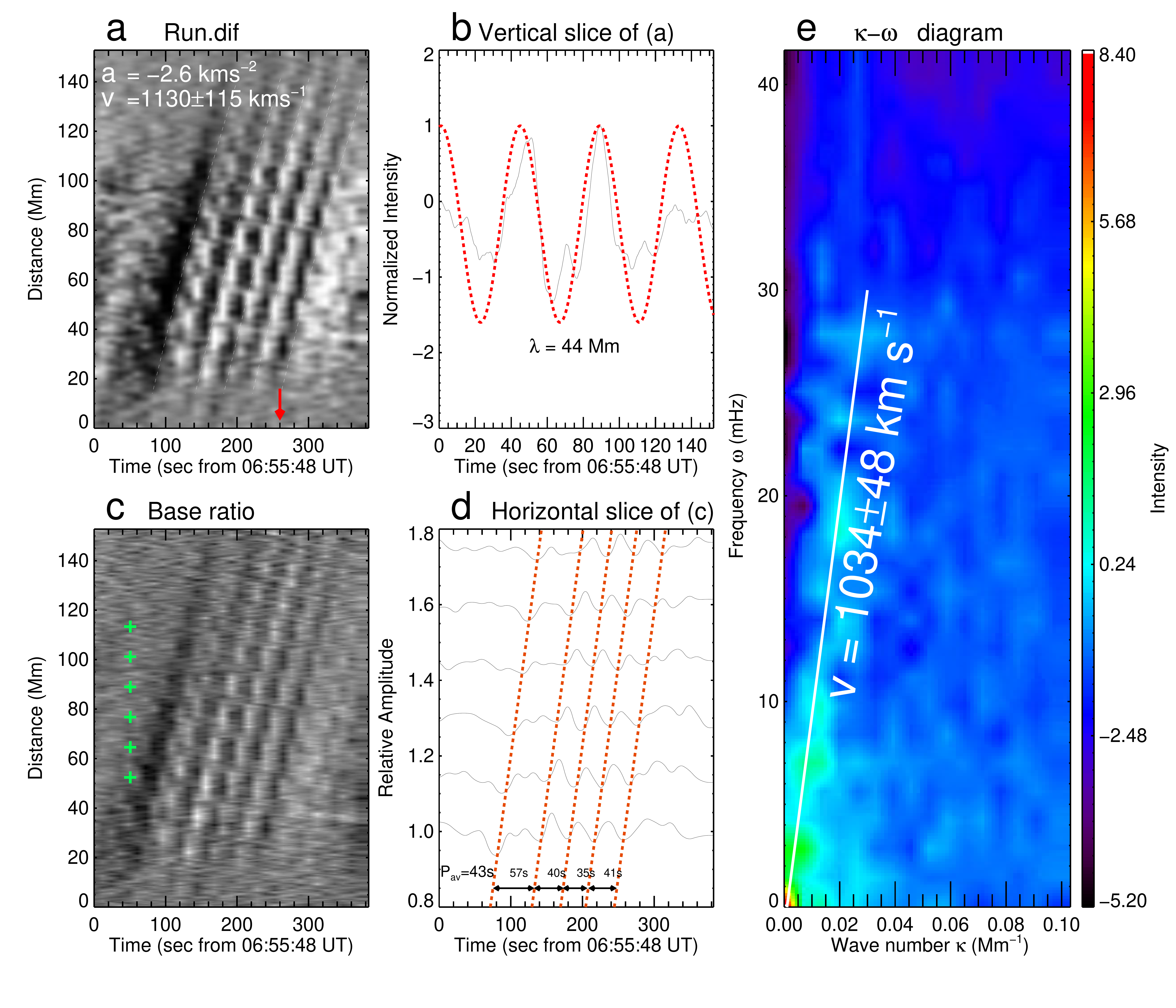}
\caption{Panel (a) is an enlargement of the wave Train-2 shown in \mfig{fig:tdp}. Panel (b) shows the wavefront profile at times marked by the red arrow in panel (a). The red dotted curve results from fitting the wave profile using the harmonic function, which shows the wavelength of 44 Mm. Panel (c) is the same as panel (a) but for base-ratio images. Panel (d) displays the horizontal slices at the selected distance in panel (c), i.e., temporal profiles of intensity base ratio at locations marked by the green cross signs. Successive curves at greater distances are shifted upward, and the periods between two slanted lines used to trace the troughs of the wavefronts are listed in the Figure. Panel (e) shows the $\kappa$\,--\,$\omega$ diagrams obtained from the Fourier analysis of a three-dimensional data cube of 171 \AA\ running difference images during 6:50:00 UT-07:03:48 UT, whose spatial dimension is indicated by the { white box in \mfig{fig:overview}(d)}. { The uncertainties in panels (a) and (e) are estimated by making the fit ten times}.\label{fig:fourier}}
\end{figure}

Since the wave signal of Train-2 was more evident and strong, we mainly used it to investigate the kinematics characters of the QFP wave trains. \nfig{fig:fourier} (a) shows the evolutional process of wave Train-2 in the running-difference images. Using the quadratic to fit the brightened ridges, we got that the wave train’s mean speed and deceleration were about 1130$\pm$\speed{115} and {\acc{2.6}}, respectively. \nfig{fig:fourier} (b) shows the evolution pattern of the wave train extracted  at 07:00 UT marked with the red arrow in \mfig{fig:fourier} (a), in which the red dotted curve is the corresponding fitting result with a sine function, indicating that the wavelength $\lambda$ was about 44 Mm. \nfig{fig:fourier} (c) is a TDSP constructed along the same slice as \mfig{fig:fourier} (a), but using the base-ratio AIA 171 \AA\ images, used to obtain the amplitude information of the wave Train-2. \nfig{fig:fourier} (d) displays the wave train’s profiles at distances marked with green crosses in \mfig{fig:fourier} (c). From the intensity profiles of the wave at different distances displayed in \mfig{fig:fourier} (d), one can find that the interval between two troughs is 57 seconds, 40 seconds, 35 seconds, and 41 seconds, indicating the average period of wave Train-2 was about 43 seconds. Meanwhile, the wave train’s max relative intensity was about $3\%-5\%$. Using the same method, we get the mean speeds (relative intensity) of Train-1 { and} Train-3 were in the range of 1035-\speed{1180} (2\%-6\%).

Following the method of \cite{2004ApJ...617L..89D}, \cite{2011ApJ...736L..13L}, and \cite{2014ApJ...797...37L}, Fourier analysis is applied to the Train-2. We extracted a three-dimensional data cube $(x, y, t)$ from a time series of 171 \AA\ running difference images for the FOV marked by the white box in \mfig{fig:overview} (d) during 06:50:00-07:03:48 UT. We then convert the data cube to the wavenumber-frequency ($\kappa$--$\omega$) diagrams via a three-dimensional Fourier transformation (as shown in \mfig{fig:fourier} (e)). A slanted ridge, representing the dispersion relation of the wave train, can be identified clearly in the $\kappa$-- $\omega$ diagram. This steep ridge can be fitted with a straight line passing through the origin point, which gives an average { phase speed ($v_{{ph}}=\omega/\kappa$) and group speed ($v_{{gr}}=d\omega/d\kappa$)} of  1034$\pm$\speed{48}. This result indicates that the wave train was non-dispersion. At the same time, the wavelength $\lambda$ = 44 Mm gotten from the sinusoidal fits (\mfig{fig:fourier} (b)) and the period $P=43$ seconds estimated above gives the { $v_{{ph}}=\lambda/P=$\speed{1046}}, which is consistent { with those} obtained from the TDSP (\mfig{fig:fourier} (b)) and $\kappa-\omega$ diagram (\mfig{fig:fourier} (e)). This result further shows that the observed wave trains were non-dispersive in their corresponding AIA channel formation heights. Similar to the narrow QFP wave propagating along the  loop system \citep{2011ApJ...736L..13L} and the broad QFP wave train running in the solar corona \citep{2022ApJ...930L...5Z}.

\begin{figure}
\centering
\plotone{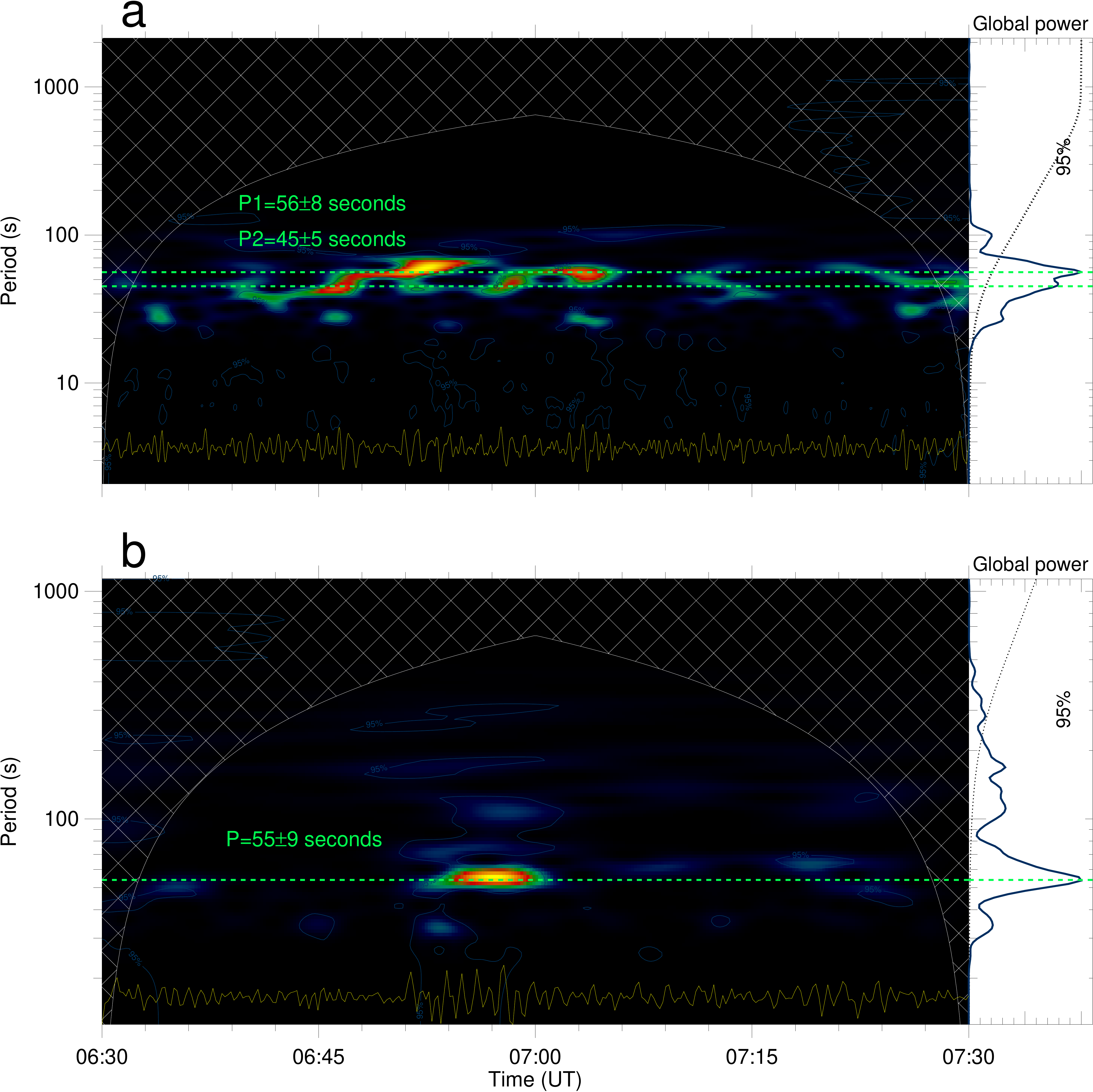}
\caption{Periodicity analysis for the associated flare pulsation and the narrow QFP wave train. The yellow curve in panel (a) is the { time derivative} signal of GOES 1-8 \AA\ X-ray flux after subtracting its 120 seconds smoothed intensity profile, while the yellow curve in panel (b) is the smoothed intensity profile along the horizontal white dashed line shown in \mfig{fig:tdp} (b). Their corresponding wavelet spectrums are displayed in panels (a) and (b). In each spectrum, the dotted line in global power indicates the 95\% significance level, and the green horizontal dashed line marks the period. The corresponding periods P are also listed in the Figure. { The uncertainty of each period is estimated by the FWHM of each peak on the global power curve}. The cross-hatched areas overlaid on the wavelet spectrum indicate the cone influence region due to the edge effect of the data.
\label{fig:wavelet}}
\end{figure}

To further examine the relationship between the flare and the wave trains, we analyze their periods using the Morelet wavelet technique \citep{1998BAMS...79...61T}. As shown in \mfig{fig:wavelet} (a), the wavelet power spectrum of the { time derivative} curve of GOES 1-8 \AA\ reveals the presence of a series of periodicity signals with periods between 45$\pm$5 and 56$\pm$8 seconds throughout the flare pulsation phase. Their distribution in time is consistent with that of the bumps in the { time derivative} curve of GOES 1-8 \AA\ in \mfig{fig:tdp} (a). Since the wave signals of Train-1 and Train-3 were relatively weak, the signal of Train-2, with a period of 55$\pm$9 seconds from 6:54 UT to 7:00 UT, dominates the wavelet power spectrum, which is different from that of the distributions of the { time derivative} curve of GOES 1-8 \AA\ in the wavelet spectrum. The similar periods between the pulsation flare and the QFP wave trains indicate that these two phenomena are possibly excited by the same physical mechanism. The periods are estimated based on the global power of the wavelet spectrum. { The period in  \mfig{fig:fourier} was obtained by fitting a finite number of brighter ridges, while the wavelet analysis analyzed the full fluctuation signal. Therefore the two have some differences. The periods obtained by both approaches are essentially similar when the uncertainties of wavelet analysis are taken into account.}

\begin{figure}
\centering
\plotone{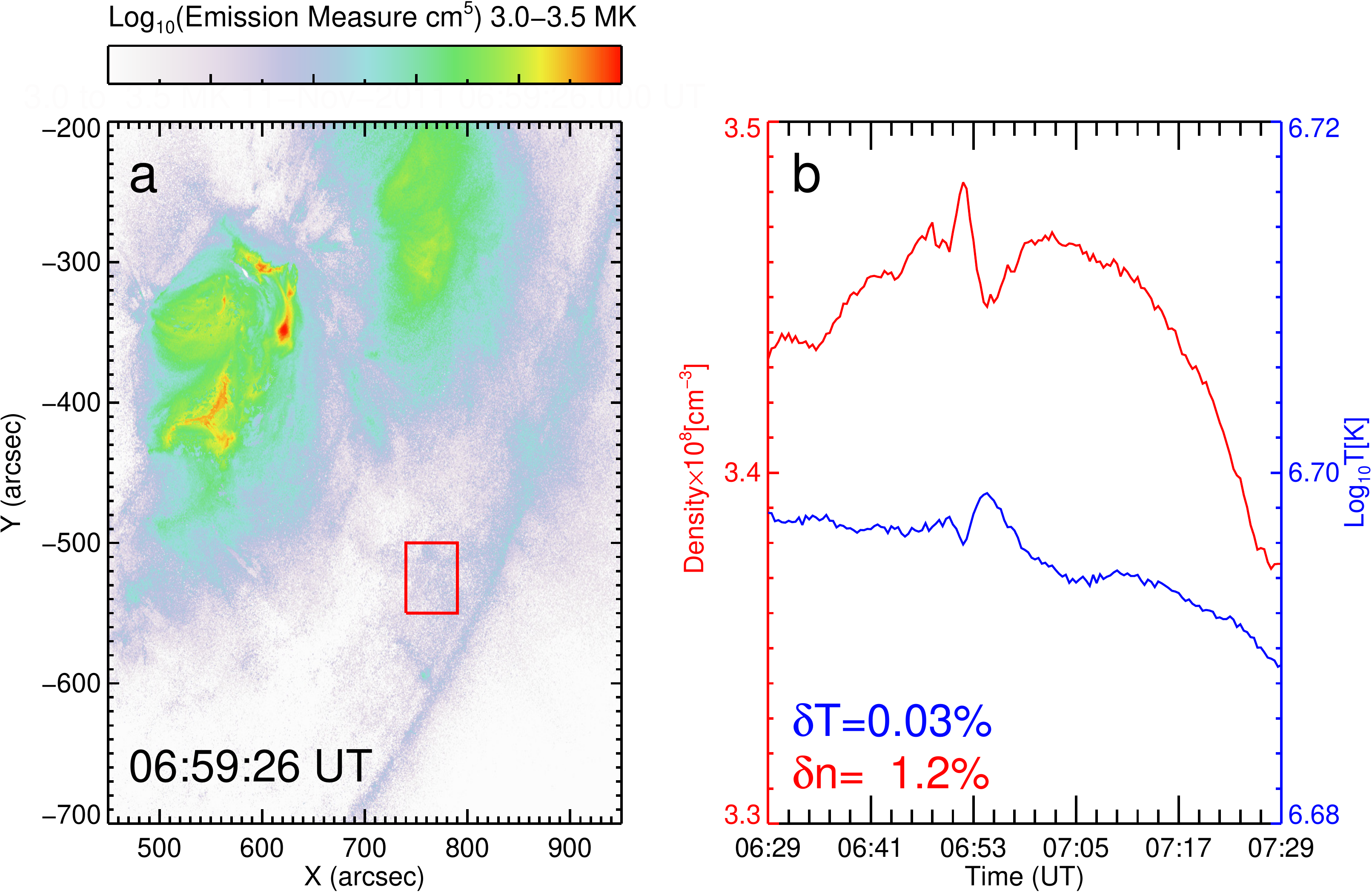}
\caption{Panel (a) shows the average temperature in the 3.0-3.5 MK using the regularized inversion technique developed by \cite{2012A&A...539A.146H}. Panel (b) displays the variation in temperature and density in the region marked by the red rectangle in panel (a) using Equations \ref{eq:t} and \ref{eq:n}. The percentage increase in density and temperature due to the wave’s passage are listed in the Figure.}
\label{fig:dem}
\end{figure}

\subsection{ Plasma Diagnostics}
Next, we calculate the Differential Emission Measure (DEM) to investigate the plasma temperature and density of the loop system during the passage of the wave train. This inversion code, developed by \cite{2012A&A...539A.146H}), combines the almost simultaneous observations of six AIA EUV channels (131 \AA, 94 \AA, 335 \AA, 211 \AA, 193 \AA, and 171 \AA), enabling an estimate to be made of the emission measure of the interested region. The DEM is defined as
\begin{equation}
{ DEM}(T)=n_{ e}^2(T)\frac{{ d}h}{{ d}T},
\end{equation}

where $n_{ e}$ is the number density dependent on the temperature $T$ along the line of sight. Following \cite{2012ApJ...761...62C, 2015ApJ...812..173V}, the average temperature and density can be defined as,
\begin{equation}
\bar{T}= { \frac{\int_{T} {   DEM}(T)T{ d}T}{\int_{T}{ DEM}(T){ d}T}}
\label{eq:t}
\end{equation}
and 
\begin{equation}
\bar{n}= \sqrt{\frac{\int_{T} { DEM}(T){ d}T}{h}},
\label{eq:n}
\end{equation}

respectively, where $h$ is the column height of emitting plasma along the line of sight taken as 8 Mm { \citep[the average value of ][]{2021ApJ...908L..37M}}. \nfig{fig:dem} (a) displays an example of a DEM map at temperature ranges from 3.0 MK to 3.5 MK at 06:59:26 UT, showing increased values in the kernel of the active region, with lower values in the quiet-Sun region. To quantify the effect of the wave trains on the plasma, we chose a small region (highlighted with a red box in \mfig{fig:dem} (a) ) in the wave propagation path to examine the temporal variation of temperature and density due to the passage of the wave. 
The DEM-derived temperature and density were estimated using Equations \ref{eq:t} and \ref{eq:n}, and the results are shown in \mfig{fig:dem} (b). Both the temperature and density of the plasma behave as expected, exhibiting an increase of 1.2\% and 0.03\% above the pre-event as a result of the passage of the wave train, respectively. Similar variations have been reported in previous works about large-scale EUV waves \citep{2019ApJ...882...90L,2021ApJ...921...61L,2022ApJ...928...98H,2022ApJ...930L...5Z}. The almost 40-fold difference between these two physical quantities suggests that the change in image intensity is more sensitive to the density variation than to the temperature \citep{2011JASTP..73.1096Z}.

\subsection{Determination of magnetic field and energy}
{ Considering the speed of the QFP wave trains was about 1130$\pm$\speed{115}, which is significantly faster than the typical sound speed ($ c_{ s}=\sqrt{\varUpsilon K_BT/m_{ p}}\approx 0.12T[K]^{1/2}\approx$ \speed{90} ) in the 0.6 MK (peak response temperature of the 171 \AA\ channel). Thus the fast-mode magnetosonic wave speed $v_f=\sqrt{v_{ A}^2+c_{ s}^2}$ is approximately equal to the Alfv\'en speed $v_{ A}$. The magnetic field strength could be written as $B \approx v_{ A}\sqrt{4\pi\rho}=v_f\sqrt{4\pi\mu m_{ p}n} $, { where $\mu$ (0.6)} is the mean molecular weight in the corona, $m_{ p}$ ($1.67\times10^{-24 }g$) is the proton mass, and $n$ is the total particle number density. According to \cite{1982soma.book.....P}, $n$ is often taken as $1.92n_{ e}$. The $n_{ e}$ is taken here as around $3.44\times10^8 cm^{-3}$, derived from DEM before the QFP wave arrival. Given that the wave trains propagated at the fast-magnetosonic speed, i.e., $v_f=v_{ ph}=$1130$\pm$\speed{115}, and taking it into the above formula, the magnetic field strength is estimated as { $B\geqslant $10\, Gauss}. This value is in the range explored by \cite{2021ApJ...908L..37M} using the extrapolation method.}

The energy flux $\mathcal{F}$ of the QFP wave trains can be calculated by employing
$\mathcal{F}=\frac{1}{2}\rho v_{ am}^2v_{ gr}\approx \frac{1}{2}\rho v_{ am}^2v_{ ph}$, where $v_{ am}$ is the disturbance amplitude of the locally perturbed plasma. % $v_{ gr}$ and $v_{ ph}$ are the wave train's group speed and phase speed, respectively.
Considering the relationship that the emission intensity is proportional to the square of the plasma density, i.e., $I \varpropto \rho^2$ in the optically thin corona, and using $v_{ am}/v_{ ph}\geqslant \delta\rho/\rho=\delta I/(2I)$, the energy flux can be rearranged as $\mathcal{F}\geqslant \frac{1}{8}\rho v_{ ph}^3(\frac{\delta I}{I})^2$ \citep{2011ApJ...736L..13L,2018ApJ...860...54O,shen2022}. If we take the phase speed and the relative amplitude of the wave trains are about $v_{ ph}=$\speed{1100} and 3\,\% (these two values are the average of Train1-3), and use the mean number density $n_{ e} \approx 3.44\times10^8  cm^{-3}$, we obtain a lower limit value of the energy flux of the wave trains in the present case, i.e., $\mathcal{F}\geqslant 1.2\,\times 10^{5}  erg\cdot cm^{-2}{ s}^{-1}$.

\section{Discussion and conclusions} 
\label{se:discussion}
In this work, we studied  QFP wave trains with a speed of about \speed{1100} { confined} in a funnel  loop system with an angle of $16^\circ$ and found that: (1) They have a tight correlation with the accompanying flare. (i) Employing the AIA images and GOES flux data, we find the onsets of the wave trains were synchronized with the energy-releasing of the flare. (ii) The wave periods were in good agreement with the observed flare.
(2). Using the Fourier analysis, we find that the wave trains have identical phase speed and group speed, indicating that the wave trains were non-dispersion waves. (3). Based on the DEM estimation, we find that both the density and the temperature of the loop system increased after the passage of wave trains, which might indicate the heating of coronal plasma due to the compression of the wave trains. In addition, we estimated the wave's energy flux ($\mathcal{F}\approx 1.2\,\times 10^{5}  erg\cdot cm^{-2}{ s}^{-1}$) and the magnetic field strength { ($B\approx$ 10\,Gauss)} of the coronal loops. 
Considering these results together, we believe that the wave trains under study should be excited by the QPP of the flare instead of attributed to the waveguide dispersion. 

Studying the narrow QFP wave trains is of particular importance in solar physics since this can provide insights into the physical nature of the disturbances as well as into the possible generation mechanisms. Generally, the generation mechanism of the narrow QFP wave train is diverse. However, the most likely relevant mechanisms are the pulsed energy release in magnetic reconnections \citep{2011ApJ...736L..13L,2012ApJ...753...53S,2013A&A...554A.144Y,2018ApJ...868L..33L,2018ApJ...855...53L} and the dispersive evolution of an { impulsive broadband perturbation} \citep{1993SoPh..144..255M,2017ApJ...847L..21P,2018MNRAS.477L...6S}. 
\begin{enumerate}

\item It is logical for the flare driving mechanism because the flare can launch a freely propagating blast wave train by generating an episodic pressure pulse intrinsic to magnetic reconnection \citep{2000SoPh..196..181V,2008SoPh..253..215V}. Thus the period is always similar to that of the flare \citep{2013SoPh..288..585S,2022ApJ...930L...5Z,2022A&A...659A.164Z,2021SoPh..296..169Z}. Observations in this study support this suggestion since that (i) the start time of the wave train were found to correlate in beginning time with the flare energy releasing, (ii) the wavelet analysis shows that the wave trains and the flare share common periods of about 45-56 seconds. Meanwhile, the wave's intensity (amplitude) is determined by the released energy of the flare; that is, the wave excited in the impulsive phase usually has a { stronger} intensity, which is also reflected in the present case.

\item The narrow QFP wave train is always constrained to propagate along open or closed coronal loops, indicating the presence of waveguides. Fast magnetoacoustic waves undergo geometric dispersion in the waveguide, which leads to localized perturbations developing into QFP wave trains, as pointed out by \cite{1983Natur.305..688R,1984ApJ...279..857R}. Numerical simulations performed by \cite{2013A&A...560A..97P,2014A&A...568A..20P} sustain this perspective. A distinctive feature of these guided waves is that they will exhibit decreasing periods, resulting in the characteristic ``tadpole'' wavelet spectral signature \citep{2004MNRAS.349..705N,2005SSRv..121..115N}, where a narrowband tail precedes a broadband head. However, the present study did not identify the expected dispersive evolution with decreasing periods. In the $\kappa$\,--\,$\omega$ diagrams, the dispersion relations of the wave trains are exhibited as bright, nearly straight ridges passing through the origin, which demonstrates that the propagation of the wave train has the same phase and group speed with a value of about 1034$\pm$\speed{48}. These two speeds were consistent with the phase speed obtained from TDSP (1130$\pm$\speed{115}) and the formula $v_{ ph}=\lambda /P$ (\speed {1046}). Therefore, we speculate that these wave trains should be excited by the intermittent energy releases rather than the dispersive mechanism. In addition, the leakage of pressure-driven $p-$modes oscillation { from the photosphere and chromosphere} into the corona \citep{2012ApJ...753...53S}, slow-mode magnetosonic waves leaking from the 3-minute chromospheric sunspot oscillations \citep{2009A&A...505..791S}, the magnetic reconnection between the loop system \citep{2017ApJ...844..149K,2018ApJ...868L..33L,shen2022}, the eruption of the magnetic flux rope \citep{2018ApJ...853....1S,2020ApJ...894...30W}, the successive stretching of magnetic field structrues \citep{2018ApJ...860L...8S,2022ApJ...939L..18S}, as well as the jet \citep{2018ApJ...861..105S,2022ApJ...926L..39D} are both { thought to be} candidate drivers of QFP wave trains.
\end{enumerate}

The source region was located at $\approx$ S20$^\circ$, and the wave propagated { toward the southwest}. As a result, it is necessary to account for the influence of foreshortening effects, { and thus the column height of emitting plasma $h$ here we taken a mean value of the previous work \citep{2021ApJ...908L..37M}, where the active region in their study was located near the center of the solar disk. However, the uncertainty remain can not be ignore, after all, the used $h$ is not the real value of the current event.} 
It is true that both the { density and temperature increase} when the wave trains passage, { although lower} than that of the broad QFP wave train in previous observation \citep[cf:][]{2022ApJ...930L...5Z}. In particular, the density increase first, followed by temperature \citep[see also][]{2019ApJ...882...90L}, which indicates that it needs some time for adiabatic compression to cause temperature rise \citep{2011ApJ...738..167S,2012ApJ...750..134D}. After the wave trains' passage, the plasma's temperature and density gradually decrease, which may be caused by the thinning of the magnetic field lines in the funnel due to the wave perturbation.

{ Based on a survey by \citep{2022SoPh..297...20S}, we find that the energy flux carried by narrow QFP wave trains (0.1-4$\,\times 10^{5}  erg\cdot cm^{-2}{ s}^{-1}$ ) is significantly smaller than that of the broad QFP wave trains (10-19$\,\times 10^{5}  erg\cdot cm^{-2}{ s}^{-1}$). The energy flux, about $ 1.2\,\times 10^{5}erg\cdot cm^{-2}{ s}^{-1}$, of current wave trains is a typical value of the narrow QFP wave train. Generally, the energy densities of the waves depend on their speeds in the respective media and the incident angle, among other quantities \citep{2014SoPh..289.3233L}. Both the narrow and broad QFP wave trains are tightly associated with the flare. However, the energy level of flares does not seem to be the critical physical condition for determining to excite which kind of wave trains \citep[see Table 1 of ][]{2022SoPh..297...20S}. Therefore, the underlying cause of the difference in energy flux density between broad and narrow waves remains an open question. Namely, the question of which physical conditions excite broad and which excite narrow waves is still unanswered.}

To conclude, the excellent spatial and temporal resolution of the SDO data presented here allows us to understand better the physical properties of the narrow QFP wave train, which are important for mode inputs and constraints. The origin of the narrow QFP wave trains remains subtle. Therefore, more numerical simulations and observational examples are required to fully { understand} the wave generation, evolution, and effect on the coronal plasma.
\section{Acknowledgements}
The authors thank the teams of SDO and GOES for providing the excellent data and Prof. Dr. Rui Liu for helping with data processing and providing valuable suggestions. The authors thank the anonymous referee for his/her valuable comments and suggestions to improve the quality of this paper. 
This work was supported by the Natural Science Foundation of China (12173083, 11922307, 11773068, 11503082), the Yunnan Science Foundation for Distinguished Young Scholars (202101AV070004), the National Key R\&D Program of China (2019YFA0405000), the Specialized Research Fund for State Key Laboratories, and the Yunnan Key Laboratory of Solar Physics and Space Science (202205AG070009), Joint Funds of the National Natural Science Foundation of China (U1931116).
%\bibliographystyle{aasjournal}{}
%\bibliography{euv_wave}

\vspace{5mm}
\end{document}